\documentclass[aps, prb, twocolumn, amsmath, amssymb, showpacs,
superscriptaddress]{revtex4-1}

\usepackage[utf8]{inputenc}
\usepackage{graphicx}
\usepackage{units}
\usepackage{color}
\usepackage[pdftex, colorlinks=true, linkcolor=myblue, citecolor=myblue,
urlcolor=myblue]{hyperref}

\definecolor{myblue}{rgb}{0,0,1}

\begin{document}

\title{Parametric amplification of magnetoplasmons in semiconductor quantum dots}

\author{Guillaume Weick} 
\email{Guillaume.Weick@ipcms.unistra.fr}
\affiliation{Institut de Physique et Chimie des Mat\'eriaux de Strasbourg (UMR
7504), CNRS and Universit\'e de Strasbourg, 23 rue du Loess, BP 43, F-67034
Strasbourg Cedex 2, France}
\author{Eros Mariani}
\affiliation{School of Physics \& Centre for Graphene Science, 
University of Exeter, Stocker Road, Exeter, EX4 4QL, United Kingdom}


\begin{abstract}
We show that the magnetoplasmon collective modes 
in quasi-two-dimensional semiconductor quantum dots 
can be parametrically amplified by periodically
modulating the magnetic field perpendicular to the nanostructure. The two
magnetoplasmon modes are excited and amplified simultaneously, leading to 
an exponential growth of the number of bosonic excitations in the system.
We further demonstrate that damping mechanisms as well as anharmonicities in the
confinement of the quantum dot lead to a saturation of the parametric
amplification. 
This work constitutes a first step towards parametric amplification of collective
modes in many-body fermionic systems beyond one dimension.
\end{abstract}


\maketitle

\section{Introduction}
The swing is the paradigm for the phenomenon of parametric resonance: \cite{landau} 
The periodic motion of the swinger's legs leads to a periodic modulation of the
effective length of the pendulum, hence of its frequency. 
This results in the exponential amplification of the motion
if the period of the modulation is chosen to be commensurate with the natural frequency
of the pendulum. 
Quantum mechanically, if one considers a single harmonic oscillator
whose frequency is periodically varied in time, the parametric modulation of the
Hamiltonian leads to an exponential divergence of the corresponding lowering and
raising operators, hence increasing the number of bosonic quanta in the system.
As in the classical case, \cite{landau} this effect is most efficient if the 
pumping frequency is twice as much as the natural frequency of the oscillator.
For a parametrically modulated many-body bosonic system, only the eigenmodes
fulfilling the resonance condition are amplified. \cite{tozzo05_PRA, goren07_PRA}
Parametric resonance is thus an interesting and valuable spectroscopic tool for many-body quantum
systems, especially in the context of cold atomic gases \cite{bloch08_RMP} where
the tunability of the experimental setups enables one to parametrically amplify
Bose-Einstein condensates. \cite{stofe04_PRL, engel07_PRL}

In this paper, we address the question of parametric resonances in many-body
\textit{fermionic} systems. In contrast to the bosonic case, any
mean-field treatment of the interactions between fermions preserves
the fermionic nature of the quasiparticles: The parametric amplification 
of fermionic modes is thus blocked due to the Pauli principle.
However, many-body fermionic systems are known to exhibit
bosonic collective modes such as, e.g., plasmons (charge density waves) and magnons (spin
waves). \cite{pines, giuliani}
It has been shown in various contexts that such collective modes can be parametrically excited
in one-dimensional correlated fermionic systems. \cite{orgad96_PRL, kagan09_PRA, graf10_EPL,
piela11_PRA} Using the Luttinger liquid picture and bosonization techniques
which enable one to treat the interactions among the particles (almost) exactly, 
\cite{giuliani, giamarchi} it was proposed that the parametric modulation of
the optical lattice where cold fermionic atoms are trapped could serve as a
probe for the well-known spin-charge separation inherent to one-dimensional
correlated fermionic systems. \cite{kagan09_PRA, graf10_EPL}

Extending these ideas to two- and three-dimensional many-body fermionic
systems is a formidable theoretical challenge, as one needs to capture the
relevant correlations by treating interactions
beyond mean field. In this paper, we focus on the case of
\textit{confined} many-body fermionic systems, where collective modes exist due
to the finite size of the system. \cite{bertsch} Examples of such collective
modes are plasmon excitations which correspond to the motion of the electronic center of mass
in metallic nanoparticles \cite{bertsch, weick06_PRB, weick11_PRB} and in quasi-two-dimensional 
semiconductor quantum dots.
\cite{jacak, reima02_RMP, mikha98_APL} 

\begin{figure}[tb]
\includegraphics[width=\linewidth]{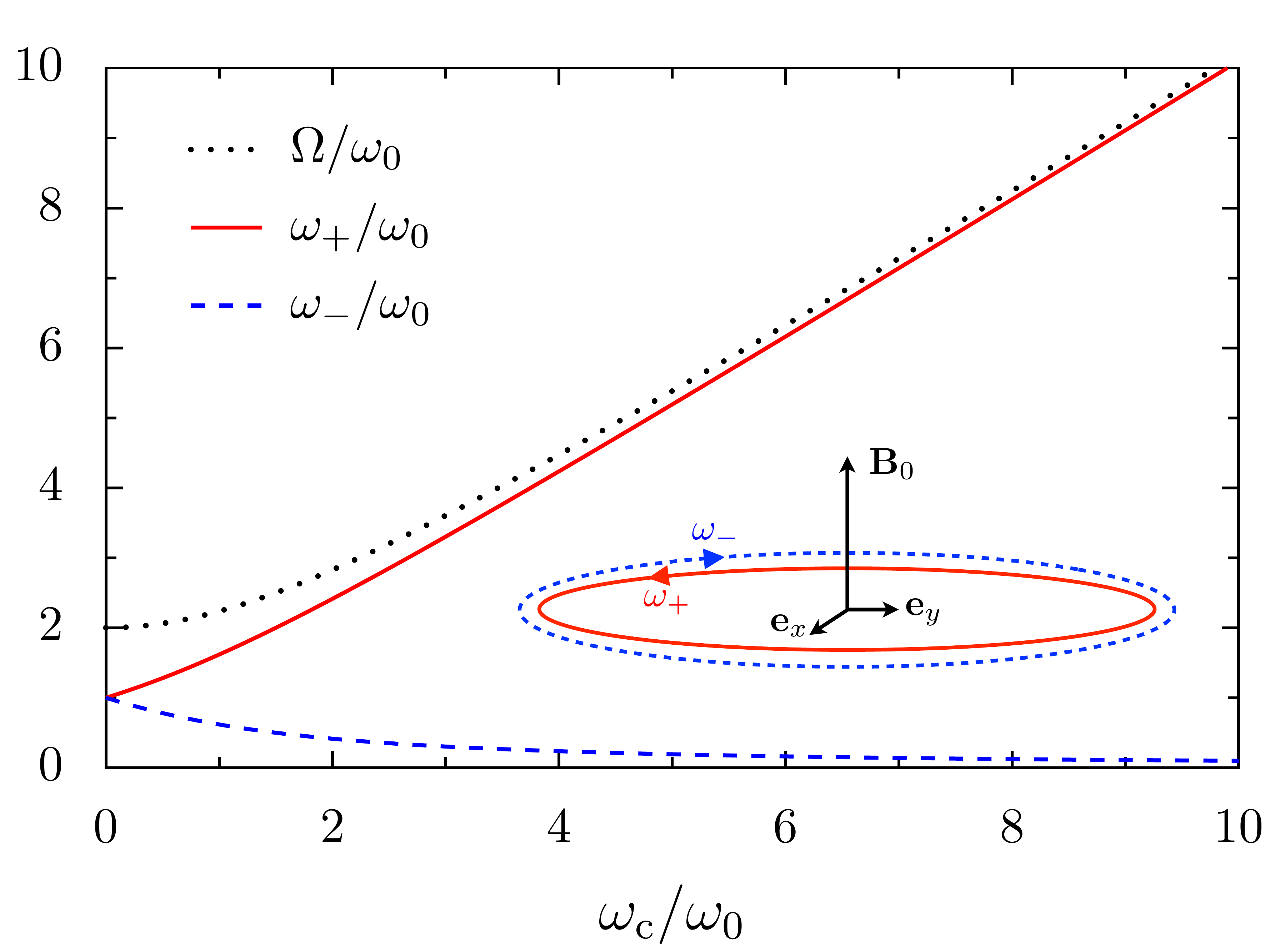}
\caption{\label{fig:spectrum}%
(Color online) 
Frequencies $\omega_+$ and $\omega_-$ [cf.\ Eq.~\eqref{eq:omega_pm}] of the two magnetoplasmon
modes (rescaled by the confinement frequency $\omega_0$) as a function of the
cyclotron frequency $\omega_\mathrm{c}=eB_0/m$. The pumping frequency exactly at
resonance, $\Omega=\omega_++\omega_-$, is also shown as a dotted line.
Inset:
Sketch of the two independent center-of-mass motions associated to the surface magnetoplasmon modes
perpendicular to the static magnetic field $\mathbf{B}_0$.}
\end{figure}

It is well known
that a static magnetic field induces field-dependent collective modes termed
magnetoplasmons. \cite{jacak, reima02_RMP} These modes appear in particular in
quasi-two-dimensional semiconductor quantum dots, \cite{allen83_PRB,
sikor89_PRL, meure92_PRL} and were recently investigated
theoretically in metallic nanoparticles. \cite{weick11_PRB} 
The magnetoplasmons, whose classical motion corresponds to an excitation
of the electronic center of mass perpendicular to the magnetic field 
rotating clockwise and counterclockwise (see the inset in Fig.\
\ref{fig:spectrum}), have distinct frequencies that depend on the applied
static magnetic field (see Fig.\ \ref{fig:spectrum}). 
In this work, we propose to trigger the parametric amplification of the
magnetoplasmons by periodically modulating the magnetic field perpendicular to
the nanostructure.
Specifically, we focus
on quasi-two-dimensional semiconductor quantum dots. 
We show that the parametric modulation of the magnetic field leads to an
exponential growth of the number of bosonic modes in the system.
We further demonstrate how damping mechanisms and
anharmonicities in the confinement of the quantum dot lead to the saturation of
the parametric amplification. 

This article is organized as follows: We start in Sec.\ \ref{sec:model} by
presenting our model of a two-dimensional quantum dot subject to a
perpendicular magnetic field which is periodically modulated in time. In Sec.\
\ref{sec:PR}, we show that such a modulation leads to the parametric
amplification of the magnetoplasmon collective modes. The effect 
of damping and anharmonicities on the parametric resonance of the magnetoplasmons
is studied in detail in Sec.\ \ref{sec:quartic}. 
We discuss in Sec.\ \ref{sec:experiments} the experimental consequences of our
proposal before we conclude 
in Sec.\ \ref{sec:ccl}. Several technical aspects of our work are presented
in the appendices.

\section{Model}
\label{sec:model}
We consider $N_\mathrm{e}$ interacting electrons with effective mass $m^*$ and charge $-e$ 
confined in a two-dimensional quantum dot. The quantum dot is subject to a
spatially homogeneous perpendicular magnetic field
$\mathbf{B}(t)=B(t)\mathbf{e}_z$ with 
\begin{equation}
\label{eq:B}
B(t)=B_0[1+\eta\sin{(\Omega t)}]
\end{equation}
which is periodically modulated at the frequency $\Omega$.
In what follows, we assume the strength of the periodic modulation
$\eta$ to be much smaller than one. The Hamiltonian of the system reads 
\begin{align}
\label{eq:H(t)_def}
H(t)=&\sum_{i=1}^{N_\mathrm{e}}\left[
\frac{1}{2m^*}\Big(\mathbf{p}_i+e\mathbf{A}(\mathbf{r}_i, t)\Big)^2
+U(r_i)
\right]\nonumber\\
&+\sum_{\substack{i,j=1\\(i\neq j)}}^{N_\mathrm{e}}
V_\textrm{e-e}(|\mathbf{r}_i-\mathbf{r}_j|), 
\end{align}
with $\mathbf{r}_i=x_i\mathbf{e}_x+y_i\mathbf{e}_y$ the position of the $i$th
electron, $\mathbf{p}_i$ its momentum and $\mathbf{A}(\mathbf{r}, t)$
the time-dependent vector potential. 
Notice that we do not consider the spin degree of freedom as it is not relevant
for understanding the parametric amplification of the magnetoplasmons, as the
latter only involves the orbital degrees of freedom. Hence, we neglect the
Zeeman term (which, for confined semiconductor quantum dots is relatively weak
due to the small value of the associated g factor) and 
the spin-orbit coupling in Eq.~\eqref{eq:H(t)_def}.
The single-particle confinement $U(r)$ appearing in Eq.\ \eqref{eq:H(t)_def} is approximated 
by a parabolic potential with confining frequency
$\omega_0$. \cite{jacak, reima02_RMP} Including also quartic corrections that
will be relevant for the following analysis (see in particular
Sec.~\ref{sec:quartic}), we write it as
\begin{equation}
\label{eq:U}
U(r)=\frac{m^*\omega_0^2}{2}r^2+\frac{a_4}{4}r^4, 
\end{equation}
with $a_4>0$.
Finally, $V_\textrm{e-e}$ in Eq.~\eqref{eq:H(t)_def} stands for the
electron-electron interaction.
In the symmetric gauge 
\begin{equation}
\label{eq:A}
\mathbf{A}(\mathbf{r},
t)=\frac{B(t)}{2}(-y\mathbf{e}_x+x\mathbf{e}_y),
\end{equation}
one can rewrite the Hamiltonian
\eqref{eq:H(t)_def} as
\begin{align}
\label{eq:H(t)}
H(t)=&\sum_i\left[\frac{p_i^2}{2m^*}+U(r_i)+\frac{\omega(t)}{2}l_{z, i}
+\frac{m^*\omega^2(t)}{8}r_i^2\right]
\nonumber\\
&+\sum_{ij}V_\textrm{e-e}(|\mathbf{r}_i-\mathbf{r}_j|), 
\end{align}
with $l_{z,i}=(\mathbf{r}_i\times\mathbf{p}_i)_z$ the $z$-component of the angular momentum, 
$\omega(t)=eB(t)/m^*=\omega_{\mathrm{c}}[1+\eta\sin{(\Omega t)]}$, 
and $\omega_{\mathrm{c}}=eB_0/m^*$ the cyclotron frequency. As we will show in 
Sec.~\ref{sec:PR}, the diamagnetic term $\propto \omega^2(t)$ in
Eq.~\eqref{eq:H(t)} is responsible for
the parametric amplification of the magnetoplasmon collective modes. 

Introducing the collective coordinate for the electronic center of mass
$\mathbf{R}=X\mathbf{e}_x+Y\mathbf{e}_y=\sum_i\mathbf{r}_i/N$ and its conjugated momentum
$\mathbf{P}=P_X\mathbf{e}_x+P_Y\mathbf{e}_y=\sum_i\mathbf{p}_i$, 
as well as the relative degrees of freedom
$\mathbf{r}_i'=\mathbf{r}_i-\mathbf{R}$ and
$\mathbf{p}_i'=\mathbf{p}_i-\mathbf{P}/N$, \cite{weick06_PRB, weick11_PRB} the Hamiltonian \eqref{eq:H(t)}
separates into
\begin{equation}
\label{eq:H(t)_separation}
H(t)=H_\mathrm{cm}(t)+H_\mathrm{rel}(t)+H_\mathrm{c}, 
\end{equation}
with the center-of-mass Hamiltonian
\begin{equation}
\label{eq:H_cm(t)_coordinates}
H_\mathrm{cm}(t)=\frac{P^2}{2M}+\frac{M}{2}\left(\omega_0^2+\frac{\omega^2(t)}{4}\right)R^2
+\frac{\omega(t)}{2}L_Z+\frac{A_4}{4}R^4, 
\end{equation}
where $M=N_\mathrm{e}m^*$, $A_4=N_\mathrm{e}a_4$, and $L_Z=XP_Y-YP_X$. The Hamiltonian for the relative
coordinates reads
\begin{align}
H_\mathrm{rel}(t)=&\sum_i\left[\frac{p_i'^2}{2m^*}+U(r_i')+\frac{\omega(t)}{2}l_{z, i}'
+\frac{m^*\omega^2(t)}{8}r_i'^2\right]
\nonumber\\
&+\sum_{ij}V_\textrm{e-e}(|\mathbf{r}'_i-\mathbf{r}'_j|). 
\end{align}
Finally, the quartic part of the single-particle confinement \eqref{eq:U} leads to the
coupling between the center of mass and the relative coordinates in Eq.\
\eqref{eq:H(t)_separation}, with 
Hamiltonian
\begin{equation}
\label{eq:H_c}
H_\mathrm{c}=\frac{a_4}{4}\sum_i\left(\left|\mathbf{R}+\mathbf{r}_i'\right|^4-R^4-r_i'^4\right). 
\end{equation}
Notice that if one assumes the single-particle confinement \eqref{eq:U}
to be purely harmonic ($a_4=0$),
there is no coupling in Eq.~\eqref{eq:H(t)_separation}
between the electronic center of mass and the relative
coordinates, as requested by the generalized Kohn's theorem. \cite{kohn61_PR,
jacak, reima02_RMP, footnote:Jacobi} The coupling \eqref{eq:H_c} induces the
damping of the center-of-mass excitations, and leads to a finite linewidth of the
magnetoplasmon lines in far infrared absorption spectroscopy
experiments. \cite{jacak, reima02_RMP}

The time-independent, quadratic part of the center-of-mass Hamiltonian
\eqref{eq:H_cm(t)_coordinates}
can be diagonalized by means of Fock-Darwin states in terms of two
magnetoplasmon excitations. \cite{jacak} Introducing the
variable $\xi=(X+\mathrm{i}Y)/\sqrt{2}$ and the
bosonic operators associated to the magnetoplasmons
\begin{subequations}
\label{eq:b_pm}
\begin{align}
b_+&=\frac{1}{\sqrt{2}}\left(\frac{\xi^*}{\ell}
+\ell\frac{\partial}{\partial\xi}\right), \quad
b_+^\dagger=\frac{1}{\sqrt{2}}\left(\frac{\xi}{\ell}
-\ell\frac{\partial}{\partial\xi^*}\right), \\
b_-&=\frac{1}{\sqrt{2}}\left(\frac{\xi}{\ell}
+\ell\frac{\partial}{\partial\xi^*}\right), \quad
b_-^\dagger=\frac{1}{\sqrt{2}}\left(\frac{\xi^*}{\ell}
-\ell\frac{\partial}{\partial\xi}\right),
\end{align}
\end{subequations}
with the oscillator length
\begin{equation}
\ell=\sqrt{\frac{\hbar}{M{(\omega_0^2+\omega_{\mathrm{c}}^2/4)}^{1/2}}},
\end{equation}
we rewrite Eq.~\eqref{eq:H_cm(t)_coordinates} as
\begin{align}
\label{eq:H_cm(t)}
H_\mathrm{cm}(t)=&
\sum_{\sigma=\pm}\hbar\omega_\sigma\left[1+\sigma\delta
f(t)\right]\left(b_\sigma^\dagger b^{\phantom{\dagger}}_\sigma+\frac 12\right)
\nonumber\\
&+\alpha\frac{\hbar\omega_0}{4}\left(b_+^\dagger+b_-\right)^2\left(b_++b_-^\dagger\right)^2
\nonumber\\
&+\frac{\hbar\omega_\mathrm{c}}{2}\delta f(t)
\left(b_+^\dagger b_-^\dagger+b_-b_+\right).
\end{align}
In Eq.~\eqref{eq:H_cm(t)}, 
\begin{equation}
\label{eq:delta_f}
\delta
f(t)=\eta\frac{\omega_\mathrm{c}/2}{\sqrt{\omega_0^2+\omega_\mathrm{c}^2/4}}\sin{(\Omega t)}, 
\end{equation}
and we introduced the dimensionless parameter
$\alpha=A_4\ell^4/\hbar\omega_0\ll1$.
The frequencies of the two magnetoplasmon collective modes read \cite{jacak}
\begin{equation}
\label{eq:omega_pm}
\omega_\pm=\sqrt{\omega_0^2+\frac{\omega_\mathrm{c}^2}{4}}\pm\frac{\omega_\mathrm{c}}{2}
\end{equation}
and are presented in Fig.~\ref{fig:spectrum} as a function of the cyclotron
frequency $\omega_\mathrm{c}$.
Without parametric modulation ($\eta=0$), the center-of-mass motion associated
with the magnetoplasmon with frequency $\omega_+$
($\omega_-$) rotates counterclockwise (clockwise) in the plane perpendicular to
the magnetic field (see the inset in Fig.~\ref{fig:spectrum}). Notice that one
has $\omega_+>\omega_-$ due to the fact that the Lorentz force
$-N_\mathrm{e}e\mathbf{\dot R}\times\mathbf{B}_0$ increases (decreases) the effective
harmonic confinement seen by the collective mode with frequency $\omega_+$
($\omega_-$).

The last symmetry breaking term $\propto b_+^\dagger b_-^\dagger+b_+b_-$ in 
the Hamiltonian \eqref{eq:H_cm(t)} 
is triggered by the parametric modulation
of the magnetic field [see Eq.~\eqref{eq:delta_f}]. As we will show in
Sec.~\ref{sec:PR}, this term is responsible for the parametric
amplification of the two magnetoplasmon collective modes, provided that the pumping
frequency $\Omega$ is close to the resonance condition
$\Omega=\omega_++\omega_-$ (see dotted line in Fig.\ \ref{fig:spectrum}). 
The quartic corrections proportional to the parameter $\alpha$ in the
Hamiltonian \eqref{eq:H_cm(t)} represent a residual interaction between the two
bosonic modes, and lead to the saturation of the parametric amplification
(see Sec.~\ref{sec:quartic}).

In the absence of quartic corrections ($\alpha=0$), the Hamiltonian \eqref{eq:H_cm(t)} is similar to the one encountered in the
quantum theory of parametric amplifiers in quantum optics. \cite{louis61_PR,
gordo63_PR, mollo67_PR, mollo67b_PR, walls} In this context, a pump laser interacts with a
nonlinear crystal. Due to the second-order susceptibility of the
nonlinear media, a photon of the pump laser at frequency $\Omega$ splits into two
photons at frequencies $\omega_+$ and $\omega_-$, a phenomenon called parametric
down conversion. In a parametric amplifier, the two signals at frequencies
$\omega_+$ and $\omega_-$ are amplified by pumping the crystal at
$\Omega=\omega_++\omega_-$.

\section{Parametric resonance of the magnetoplasmons}
\label{sec:PR}
We start our study of the possibility of parametrically amplifying the
magnetoplasmon collective modes by neglecting the quartic part of the
center-of-mass Hamiltonian, setting $\alpha=0$ in Eq.~\eqref{eq:H_cm(t)}.
Furthermore, we neglect all other sources of dissipation on top of the one
induced by the quartic confinement. 
(For the simpler case of a single harmonic oscillator with
periodically-modulated frequency, see Appendix \ref{sec:PR_HO}.)
The Heisenberg equations of motion for the bosonic operators appearing in the
center-of-mass Hamiltonian \eqref{eq:H_cm(t)} thus read
\begin{equation}
\label{eq:eom_b_pm}
\dot b_\pm=-\mathrm{i}\omega_\pm\left[1\pm\delta f(t)\right]b_\pm
-\mathrm{i}\frac{\omega_\mathrm{c}}{2}\delta f(t)b_\mp^\dagger.
\end{equation}
Introducing
\begin{equation}
b_\pm(t)=\exp{\left(-\mathrm{i}\omega_\pm\int_0^t\mathrm{d}s\left[1\pm\delta f(s)\right]\right)}
\;\tilde b_\pm(t),
\end{equation}
Eq.~\eqref{eq:eom_b_pm} transforms, to first order in the small parameter
$\eta$, into
\begin{equation}
\dot{\tilde b}_\pm=-\mathrm{i}\frac{\omega_\mathrm{c}}{2}\delta f(t)\,
\mathrm{e}^{\mathrm{i}(\omega_++\omega_-)t}\,\tilde b_\mp^\dagger.
\end{equation}
The main contribution to the time evolution of these operators, within 
the rotating wave approximation, comes from
the terms close to resonance, i.e., $\Omega\simeq\omega_++\omega_-$ [cf.\
Eq.~\eqref{eq:delta_f}], yielding 
\begin{equation}
\label{eq:system}
\dot{\tilde b}_\pm\simeq\frac{\epsilon}{2}\,
\mathrm{e}^{-\mathrm{i}(\Omega-\omega_+-\omega_-)t}\,\tilde b_\mp^\dagger,
\end{equation}
where we introduced the ``driving" frequency
\begin{equation}
\label{eq:epsilon}
\epsilon=\eta\frac{(\omega_\mathrm{c}/2)^2}{\sqrt{\omega_0^2+\omega_\mathrm{c}^2/4}}.
\end{equation}

Equation~\eqref{eq:system} can be easily decoupled, yielding
\begin{equation}
\ddot{\tilde b}_\pm+\mathrm{i}\left(\Omega-\omega_+-\omega_-\right)\dot{\tilde b}_\pm
-\frac{\epsilon^2}{4}\tilde b_\pm=0.
\end{equation}
Using the initial conditions $\tilde b_\pm(0)=b_\pm(0)$ and 
$\dot{\tilde b}_\pm(0)=\epsilon b_\mp^\dagger(0)/2$ [cf.\ Eq.\ \eqref{eq:system}], 
we finally obtain the solutions
\begin{align}
\label{eq:full_solution}
b_\pm(t)=&\;\frac{\mathrm{e}^{-\mathrm{i}\omega_\pm t}}{\Omega_--\Omega_+}
\bigg[
\left(\Omega_-\mathrm{e}^{\mathrm{i}\Omega_+t}-\Omega_+\mathrm{e}^{\mathrm{i}\Omega_-t}\right)b_\pm(0)
\nonumber\\
&+\mathrm{i}\frac{\epsilon}{2}
\left(\mathrm{e}^{\mathrm{i}\Omega_+t}-\mathrm{e}^{\mathrm{i}\Omega_-t}\right)b_\mp^\dagger(0)
\bigg],
\end{align}
with 
\begin{equation}
\label{eq:Omega_pm}
\Omega_\pm=\frac{\omega_++\omega_--\Omega}{2}\pm\frac 12
\sqrt{(\Omega-\omega_+-\omega_-)^2-\epsilon^2}.
\end{equation}
We thus see that tuning the pumping frequency $\Omega$ in the interval
$|\Omega-\omega_+-\omega_-|<\epsilon$, the frequencies $\Omega_\pm$ acquire an
imaginary part, leading to the parametric amplification of
the bosonic magnetoplasmon collective modes. 
(For a classical treatment of the parametric resonance of the magnetoplasmons
and specifically of the associated classical trajectories of the electronic
center of mass, see Appendix \ref{sec:PR_classical}.)
Indeed, close to resonance, i.e., for $|\Omega-\omega_+-\omega_-|\ll\epsilon$,
Eq.~\eqref{eq:full_solution} simplifies to
\begin{equation}
\label{eq:b_at_resonance}
b_\pm(t)=\mathrm{e}^{-\mathrm{i}\omega_\pm t}\left[\cosh{\left(\frac{\epsilon
t}{2}\right)}b_\pm(0)
+\sinh{\left(\frac{\epsilon t}{2}\right)}b_\mp^\dagger(0)\right],
\end{equation}
showing that the two bosonic modes are exponentially amplified at a rate given
by the frequency $\epsilon$ of Eq.~\eqref{eq:epsilon}.
On the contrary, off resonance ($|\Omega-\omega_+-\omega_-|\gg\epsilon$), the bosonic
operators evolve in time according to their unitary evolution, 
\begin{equation}
b_\pm(t)=\mathrm{e}^{-\mathrm{i}\omega_\pm t}b_\pm(0).
\end{equation}
Notice that the exponential amplification of the operators
\eqref{eq:full_solution} is a direct consequence of their bosonic nature. If
the operators $b_\pm$ were fermionic, the Pauli principle would prevent the parametric amplification
from occurring (see Appendix \ref{sec:toy} for details).

The center-of-mass subsystem is, prior to the parametric modulation that starts
at $t=0$, in thermal equilibrium at the temperature $T$ and thus described by the
density matrix
$\rho_0=\mathrm{e}^{-H_\mathrm{cm}(0)/k_\mathrm{B}T}/\mathcal{Z}$ with
$\mathcal{Z}=\mathrm{tr}\{\mathrm{e}^{-H_\mathrm{cm}(0)/k_\mathrm{B}T}\}$ the
canonical partition function.
Introducing the average number of bosons in each mode
\begin{equation}
\label{eq:N_pm_def}
N_\pm(t)=\langle b_\pm^\dagger b^{\phantom{\dagger}}_\pm\rangle(t),
\end{equation}
as well as
the ``anomalous" average 
\begin{equation}
\label{eq:phi_def}
\varphi(t)=\langle b_+b_-\rangle(t)
\end{equation}
where $\langle\mathcal{O}\rangle(t)=\mathrm{tr}\{\rho_0\mathcal{O}(t)\}$, 
we obtain using Eq.~\eqref{eq:b_at_resonance}
\begin{align}
\label{eq:N_pm}
N_\pm(t)=&\;\frac{1}{2}
\big\{\left[1+n_\mathrm{B}(\omega_+)+n_\mathrm{B}(\omega_-)\right]
\cosh{(\epsilon t)}
\nonumber\\
&+n_\mathrm{B}(\omega_\pm)-1-n_\mathrm{B}(\omega_\mp)
\big\}
\end{align}
and 
\begin{equation}
\label{eq:phi}
\varphi(t)=\frac 12 \mathrm{e}^{-\mathrm{i}\Omega t}
\left[1+n_\mathrm{B}(\omega_+)+n_\mathrm{B}(\omega_-)\right]
\sinh{(\epsilon t)}
\end{equation}
when the pumping frequency is close to resonance. In Eqs.~\eqref{eq:N_pm} and
\eqref{eq:phi}, we introduced the Bose distribution
$n_\mathrm{B}(\omega)=[\exp{(\hbar\omega/k_\mathrm{B}T)}-1]^{-1}$.
Due to the fact that the magnetoplasmons are purely harmonic excitations of the
electronic center of mass,  $\langle R\rangle(t)=0$, the fluctuations of the
center-of-mass coordinate $\langle \delta R^2\rangle(t)=\langle
R^2\rangle(t)-\langle R\rangle^2(t)$ reduce to
\begin{equation}
\label{eq:R2_def}
\langle \delta R^2\rangle(t)=\ell^2 
[N_+(t)+N_-(t)+1+\varphi(t)+\varphi^*(t)].
\end{equation}
With Eqs.~\eqref{eq:N_pm} and \eqref{eq:phi}, we obtain
\begin{align}
\langle \delta R^2\rangle(t)=&\;\ell^2\left[1+n_\mathrm{B}(\omega_+)+n_\mathrm{B}(\omega_-)\right]
\nonumber\\
&\times\left[\cosh{(\epsilon t)}+\cos{(\Omega t)}\sinh{(\epsilon t)}\right].
\end{align}
Since $\epsilon\ll\Omega$, one can average this equation over a timescale long
compared to $\Omega^{-1}$, but short compared to $\epsilon^{-1}$, yielding
\begin{equation}
\label{eq:R2_simple}
\overline{\langle \delta R^2\rangle}(t)=\ell^2\left[1+n_\mathrm{B}(\omega_+)+n_\mathrm{B}(\omega_-)\right]
\cosh{(\epsilon t)}.
\end{equation}
The fluctuations of the center-of-mass are thus exponentially amplified due to
the parametric modulation of the magnetic field. 
Notice that even at zero temperature, where
initially the two bosonic modes are unoccupied [$n_\mathrm{B}(\omega_\pm)=0$ in
Eq.~\eqref{eq:R2_simple}], parametric amplification takes place, triggered by
the quantum fluctuations of the center-of-mass coordinate.

\section{Effects of damping and anharmonicities on the parametric resonance of
the magnetoplasmons}
\label{sec:quartic}
In this section, we analyze the role played by dissipation and  non-linearities on the
parametric amplification of the magnetoplasmons.
As we will show, these two effects are indeed limiting the exponential amplification of the
bosonic modes presented in Sec.~\ref{sec:PR}. It is therefore relevant to study
them as they will inevitably affect experiments.

\subsection{Mean-field equations of motion}
In what follows, we adopt a different strategy than the one used in Sec.~\ref{sec:PR},
and focus on the time evolution of the averaged quantities \eqref{eq:N_pm_def}
and \eqref{eq:phi_def} as we are primarily interested in the 
fluctuations of the center-of-mass coordinate, Eq.~\eqref{eq:R2_def}.
To this end, we use the time evolution of the reduced density matrix $\rho$ associated
to the center-of-mass degrees of freedom, 
\begin{equation}
\label{eq:rho}
\dot \rho=-\frac{\mathrm{i}}{\hbar}\left[H_\mathrm{cm}, \rho\right]
+\sum_{\sigma=\pm}\frac{\gamma_\sigma}{2}
\left(2b^{\phantom{\dagger}}_\sigma\rho b_\sigma^\dagger
-b_\sigma^\dagger b^{\phantom{\dagger}}_\sigma\rho
-\rho b_\sigma^\dagger b^{\phantom{\dagger}}_\sigma\right).
\end{equation}
The first term on the right-hand side of Eq.~\eqref{eq:rho} accounts for the
unitary dynamics of the center-of-mass density matrix that evolves according to the
Hamiltonian \eqref{eq:H_cm(t)}. The second term
accounts for dissipative processes coming from the anharmonicites [and hence
from the coupling Hamiltonian \eqref{eq:H_c}], as well as stemming from other
sources of dissipation, such as radiative damping, 
interaction with phonons, Joule heating due to the eddy currents generated by
the electric field associated with the time-dependent vector potential
\eqref{eq:A}, etc. In Eq.~\eqref{eq:rho}, we take dissipation phenomenologically into
account by assuming the Lindblad form for the reduced density matrix. 
\cite{breuer} We denote $\gamma_+$ and $\gamma_-$ the damping rates for the two
magnetoplasmon modes. Notice that, in general,
$\gamma_+\neq\gamma_-$ due to the difference in energy of the two modes, as
recently demonstrated in the context of the Landau damping of the magnetoplasmon excitations in metallic
nanoparticles. \cite{weick11_PRB}

From the master equation \eqref{eq:rho}, the
time evolution of the quantum average of an operator $\mathcal{O}$ is easily
determined using the identity
$\langle\dot{\mathcal{O}}\rangle(t)=\mathrm{tr}\{\dot \rho(t)\mathcal{O}\}$.
Introducing the notation $N=N_++N_-$, $\Delta
N=N_+-N_-$, $\gamma=(\gamma_++\gamma_-)/2$, and
$\Delta\gamma=(\gamma_+-\gamma_-)/2$, we find from Eqs.~\eqref{eq:H_cm(t)} and \eqref{eq:rho}
\begin{subequations}
\label{eq:eom_impossible}
\begin{align}
\dot N=&\;
\mathrm{i}\omega_\mathrm{c}\delta f(t)\left(\varphi-\varphi^*\right)
-\gamma N-\Delta\gamma\Delta N
\nonumber\\
&-\mathrm{i}\alpha\omega_0
\Big[
2\left(\varphi^*
-\varphi\right)
+\langle b_+^\dagger b_+^\dagger b_-^\dagger b_-^\dagger\rangle
+\langle b_+^\dagger b_+^\dagger b_-^\dagger b^{\phantom{\dagger}}_+\rangle
\nonumber\\
&+\langle b_+^\dagger b_-^\dagger b_-^\dagger b^{\phantom{\dagger}}_-\rangle
-\langle b_+^\dagger b^{\phantom{\dagger}}_+b^{\phantom{\dagger}}_+b^{\phantom{\dagger}}_-\rangle
-\langle b_-^\dagger b^{\phantom{\dagger}}_+b^{\phantom{\dagger}}_-b^{\phantom{\dagger}}_-\rangle
\nonumber\\
&-\langle b^{\phantom{\dagger}}_+b^{\phantom{\dagger}}_+b^{\phantom{\dagger}}_-b^{\phantom{\dagger}}_-\rangle
\Big],
\\
\Delta \dot N=&-\gamma\Delta N-\Delta\gamma N,
\\
\dot\varphi=&
-\mathrm{i}\left[\omega_++\omega_-+\omega_\mathrm{c}\delta f(t)\right]\varphi
-\mathrm{i}\frac{\omega_\mathrm{c}}{2}\delta f(t)\left(N+1\right)
\nonumber\\
&-\gamma\varphi
-\mathrm{i}\frac{\alpha\omega_0}{2}
\Big(
2+6\varphi+2\varphi^*+4N
+\langle b_+^\dagger b_+^\dagger b_-^\dagger b^{\phantom{\dagger}}_+\rangle
\nonumber\\
&+\langle b_+^\dagger b_-^\dagger b_-^\dagger b^{\phantom{\dagger}}_-\rangle
+4\langle b_+^\dagger b_-^\dagger b^{\phantom{\dagger}}_+ b^{\phantom{\dagger}}_-\rangle
+\langle b_+^\dagger b_+^\dagger b^{\phantom{\dagger}}_+ b^{\phantom{\dagger}}_+\rangle
\nonumber\\
&+\langle b_-^\dagger b_-^\dagger b^{\phantom{\dagger}}_- b^{\phantom{\dagger}}_-\rangle
+3 \langle b_+^\dagger b^{\phantom{\dagger}}_+ b^{\phantom{\dagger}}_+ b^{\phantom{\dagger}}_-\rangle
+3 \langle b_-^\dagger b^{\phantom{\dagger}}_+ b^{\phantom{\dagger}}_- b^{\phantom{\dagger}}_-\rangle
\nonumber\\
&+2 \langle b^{\phantom{\dagger}}_+b^{\phantom{\dagger}}_+b^{\phantom{\dagger}}_-b^{\phantom{\dagger}}_-\rangle
\Big).
\end{align}
\end{subequations}
To first order in the small anharmonicity parameter $\alpha\ll1$, the
four-operator correlators appearing in Eq.~\eqref{eq:eom_impossible} must be
evaluated to order $\alpha^0$. Since for $\alpha=0$, the Hamiltonian of
Eq.~\eqref{eq:H_cm(t)} is quadratic in the bosonic operators, one can apply
Wick's theorem, yielding the mean-field equations of motion
\begin{subequations}
\label{eq:eom_mean_field}
\begin{align}
\dot N=&\;
\mathrm{i}\omega_\mathrm{c}\delta f(t)\left(\varphi-\varphi^*\right)
-\gamma N-\Delta\gamma\Delta N
\nonumber\\
&+2\mathrm{i}\alpha\omega_0
\left[\left(\varphi-\varphi^*\right)\left(N+1\right)+\varphi^2-\varphi^{*2}\right],
\\
\Delta \dot N=&-\gamma\Delta N-\Delta\gamma N, 
\\
\dot\varphi=&
-\mathrm{i}\left[\omega_++\omega_-+\omega_\mathrm{c}\delta f(t)\right]\varphi
-\mathrm{i}\frac{\omega_\mathrm{c}}{2}\delta f(t)\left(N+1\right)
\nonumber\\
&-\gamma\varphi
-\mathrm{i}\alpha\omega_0
\Big[
\left(N+1\right)^2+2\varphi^*\varphi+2\varphi^2
\nonumber\\
&+\left(3\varphi+\varphi^*\right)\left(N+1\right)
\Big].
\end{align}
\end{subequations}
Introducing
\begin{equation}
\varphi(t)=\exp{\left(-\mathrm{i}\int_0^t\mathrm{d}s
\left[\omega_++\omega_-+\omega_\mathrm{c}\delta
f(s)\right]\right)}\tilde\varphi(t), 
\end{equation}
to first order in $\eta\ll1$ and within the rotating wave approximation
close to resonance ($\Omega\simeq\omega_++\omega_-$),
we find from Eq.~\eqref{eq:eom_mean_field} 
\begin{subequations}
\label{eq:eom_mean_field_RWA}
\begin{align}
\dot N=&\;\epsilon\left[\mathrm{e}^{\mathrm{i}(\Omega-\omega_+-\omega_-)t}\tilde\varphi
+\mathrm{c.c.}\right]
-\gamma N-\Delta\gamma\Delta N
\nonumber\\
&-2\alpha\omega_0\frac{\epsilon}{\Omega}
\left[\mathrm{e}^{\mathrm{i}(\Omega-\omega_+-\omega_-)t}\tilde\varphi
+\mathrm{c.c.}\right]
(N+1),
\\
\Delta \dot N=&-\gamma\Delta N-\Delta\gamma N, 
\\
\dot{\tilde\varphi}=&\;\frac{\epsilon}{2}\mathrm{e}^{-\mathrm{i}(\Omega-\omega_+-\omega_-)t}
(N+1)-\gamma\tilde\varphi
-3\mathrm{i}\alpha\omega_0\tilde\varphi\left(N+1\right)
\nonumber\\
&-\alpha\omega_0\frac{\epsilon}{\Omega}
\left\{
\mathrm{e}^{-\mathrm{i}(\Omega-\omega_+-\omega_-)t}\left[\left(N+1\right)^2+2\tilde\varphi\tilde\varphi^*\right]
\right.
\nonumber\\
&\left.-2\mathrm{e}^{\mathrm{i}(\Omega-\omega_+-\omega_-)t}\tilde\varphi^2
\right\}.
\end{align}
\end{subequations}

\subsection{Effect of dissipative processes on the parametric resonance of the magnetoplasmon
excitations}
\label{sec:damping}
It is instructive to analyze the effect of damping alone on the parametric
resonance of the magnetoplasmon excitations. To this end, we
set the anharmonicity parameter $\alpha$ to zero
in Eq.~\eqref{eq:eom_mean_field_RWA}. 
Using the initial conditions
$N(0)=n_\mathrm{B}(\omega_+)+n_\mathrm{B}(\omega_-)$, 
$\Delta N(0)=n_\mathrm{B}(\omega_+)-n_\mathrm{B}(\omega_-)$, 
and $\tilde\varphi(0)=0$, we obtain the solution at resonance
($\Omega=\omega_++\omega_-$)
\begin{subequations}
\label{eq:solution_damping}
\begin{align}
\label{eq:solution_damping_N}
N(t)=&\,\frac{\mathrm{e}^{-\gamma t}}{\sqrt{\epsilon^2+\Delta\gamma^2}}h'(t)
-\frac{\epsilon^2}{\epsilon^2+\Delta\gamma^2-\gamma^2},
\\
\label{eq:solution_damping_delta_N}
\Delta N(t)=&\,
\frac{\epsilon\,\mathrm{e}^{-\gamma t}}{\sqrt{\epsilon^2+\Delta\gamma^2}}
\Bigg[
\frac{\epsilon}{\sqrt{\epsilon^2+\Delta\gamma^2}}
\left(
\Delta N(0)-\frac{\Delta\gamma}{\gamma}
\right)
\nonumber\\
&-\frac{\Delta\gamma }{\epsilon}h(t)
\Bigg]
+\frac{\Delta\gamma}{\gamma}\frac{\epsilon^2}{\epsilon^2+\Delta\gamma^2-\gamma^2},
\\
\tilde\varphi(t)=&\,
\frac{\epsilon\,\mathrm{e}^{-\gamma t}}{2\sqrt{\epsilon^2+\Delta\gamma^2}}
\Bigg[
\frac{\Delta\gamma}{\sqrt{\epsilon^2+\Delta\gamma^2}}
\left(
\Delta N(0)-\frac{\Delta\gamma}{\gamma}
\right)
\nonumber\\
&+h(t)
\Bigg]
-\frac{\epsilon}{2\gamma}\frac{\gamma^2-\Delta\gamma^2}{\epsilon^2+\Delta\gamma^2-\gamma^2},
\end{align}
\end{subequations}
where we defined the function
\begin{align}
\label{eq:h(t)}
h(t)=&\,
\left[
\frac{\epsilon^2}{\epsilon^2+\Delta\gamma^2-\gamma^2}+N(0)
\right]
\sinh{\left(\sqrt{\epsilon^2+\Delta\gamma^2}\,t\right)}
\nonumber\\
&+\left[
\frac{\gamma\epsilon^2}{\epsilon^2+\Delta\gamma^2-\gamma^2}
-\Delta \gamma\Delta N(0)
\right]
\nonumber\\
&\times
\frac{\cosh{\left(\sqrt{\epsilon^2+\Delta\gamma^2}\,t\right)}}{\sqrt{\epsilon^2+\Delta\gamma^2}}, 
\end{align}
and $h'(t)$ denotes its derivative with respect to time.

\begin{figure}[tb]
\includegraphics[width=\linewidth]{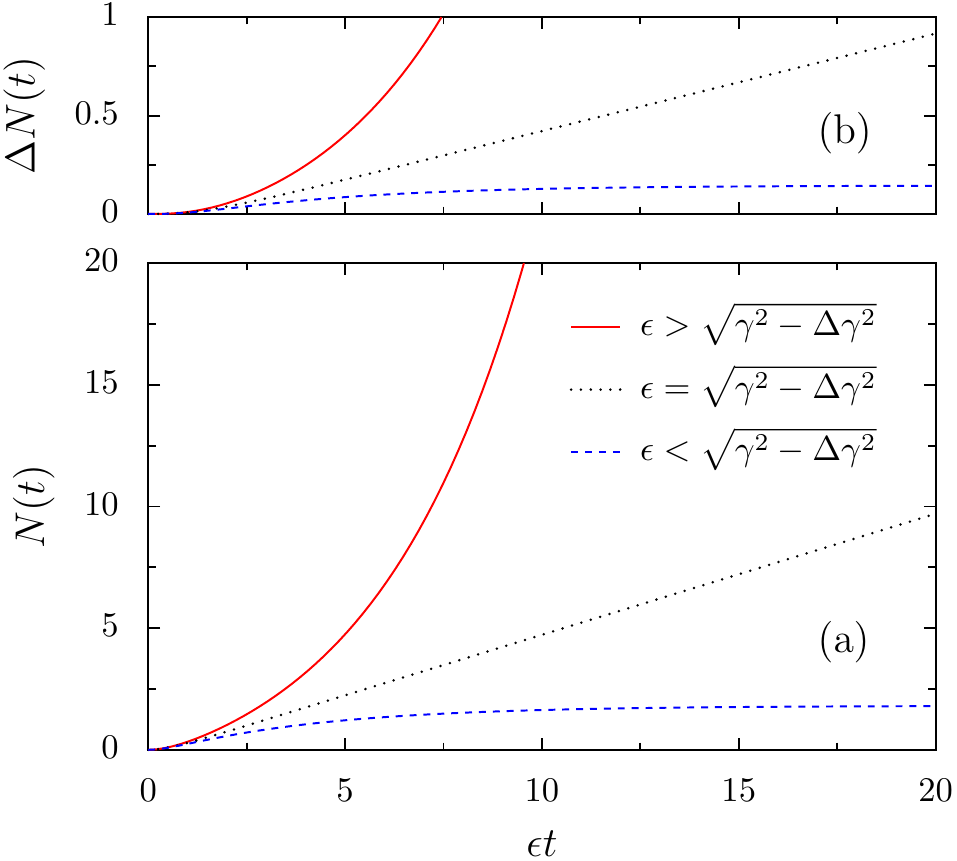}
\caption{\label{fig:damping}%
(Color online) 
(a) Total number $N(t)$ and (b) difference $\Delta N(t)$ of bosonic excitations
between the modes $+$ and $-$
as a function of time from Eqs.~\eqref{eq:solution_damping_N} and
\eqref{eq:solution_damping_delta_N}, respectively. In the figure, $T=0$,
$\Delta\gamma=-\epsilon/10$, and $\gamma=3\epsilon/4$, $\sqrt{1.01}\epsilon$,
and $5\epsilon/4$ for the solid (red), dotted (black), and dashed (blue) lines,
respectively.}
\end{figure}

Thus, for $\epsilon>\sqrt{\gamma^2-\Delta\gamma^2}$, the solutions \eqref{eq:solution_damping} 
are exponentially amplified, as shown by the solid (red) lines in
Fig.~\ref{fig:damping}.
Indeed, from
Eq.~\eqref{eq:solution_damping}, we obtain for
$t\gg(\sqrt{\epsilon^2+\Delta\gamma^2}-\gamma)^{-1}$ and $\epsilon>\sqrt{\gamma^2-\Delta\gamma^2}$
\begin{subequations}
\begin{align}
\label{eq:damping_exp_N}
N(t)\simeq&\;\frac{\mathrm{e}^{(\sqrt{\epsilon^2+\Delta\gamma^2}-\gamma)t}}{2}
\left[
N(0)-\frac{\Delta\gamma}{\sqrt{\epsilon^2+\Delta\gamma^2}}\Delta N(0)
\right.
\nonumber\\
&+\left.\frac{\epsilon^2}{\sqrt{\epsilon^2+\Delta\gamma^2}\left(\sqrt{\epsilon^2+\Delta\gamma^2}-\gamma\right)}
\right],
\\
\Delta N(t)\simeq&\;-\frac{\Delta\gamma}{\sqrt{\epsilon^2+\Delta\gamma^2}}N(t),\\
\tilde\varphi(t)\simeq&\;\frac{\epsilon}{2\sqrt{\epsilon^2+\Delta\gamma^2}}N(t).
\end{align}
\end{subequations}
On the contrary, for
$\epsilon<\sqrt{\gamma^2-\Delta\gamma^2}$, the system
reaches the stationary solution
\begin{subequations}
\begin{align}
\label{eq:N_st_damping}
N^\mathrm{st}=&\;\frac{\epsilon^2}{\gamma^2-\Delta\gamma^2-\epsilon^2},
\\
\Delta N^\mathrm{st}=&\;-\frac{\Delta\gamma}{\gamma}N^\mathrm{st},
\\
\tilde\varphi^\mathrm{st}=&\;\frac{\epsilon}{2\gamma}\left(N^\mathrm{st}+1\right), 
\end{align}
\end{subequations}
as exemplified by the dashed (blue) lines in Fig.~\ref{fig:damping}.
In terms of bosonic excitations in the two modes $+$ and $-$, this translates into
\begin{equation}
N_\pm^\mathrm{st}=\frac{\gamma_\mp}{\gamma_++\gamma_-}\frac{\epsilon^2}{\gamma_+\gamma_--\epsilon^2}, 
\end{equation}
i.e., the detailed balance $\gamma_+N_+^\mathrm{st}=\gamma_-N_-^\mathrm{st}$.
For the special case $\epsilon=\sqrt{\gamma^2-\Delta\gamma^2}$, we obtain from
Eqs.~\eqref{eq:solution_damping} and \eqref{eq:h(t)}, and for $\gamma t\gg1$ 
\begin{subequations}
\begin{align}
N(t)\simeq&\;\frac{1}{2}\left[\frac{\epsilon^2}{\gamma}\left(t-\frac{1}{2\gamma}\right)
+N(0)-\frac{\Delta\gamma}{\gamma}\Delta N(0)\right],
\\
\Delta N(t)\simeq&\;-\frac{\Delta\gamma}{2\gamma}\left[\frac{\epsilon^2}{\gamma}\left(t-\frac{3}{2\gamma}\right)
+N(0)-\frac{\Delta\gamma}{\gamma}\Delta N(0)\right],
\\
\tilde\varphi(t)\simeq&\;\frac{\epsilon}{4\gamma}\left[\frac{\epsilon^2}{\gamma}\left(t-\frac{3}{2\gamma}\right)+2
+N(0)-\frac{\Delta\gamma}{\gamma}\Delta N(0)\right],
\end{align}
\end{subequations}
i.e., a linear behavior of the solution as a function of $t$ [see dotted (black)
lines in Fig.~\ref{fig:damping}].

\subsection{Effect of anharmonicities on the parametric resonance of the magnetoplasmon
excitations}
\label{sec:anharmonic}

\begin{figure}[tb]
\includegraphics[width=\linewidth]{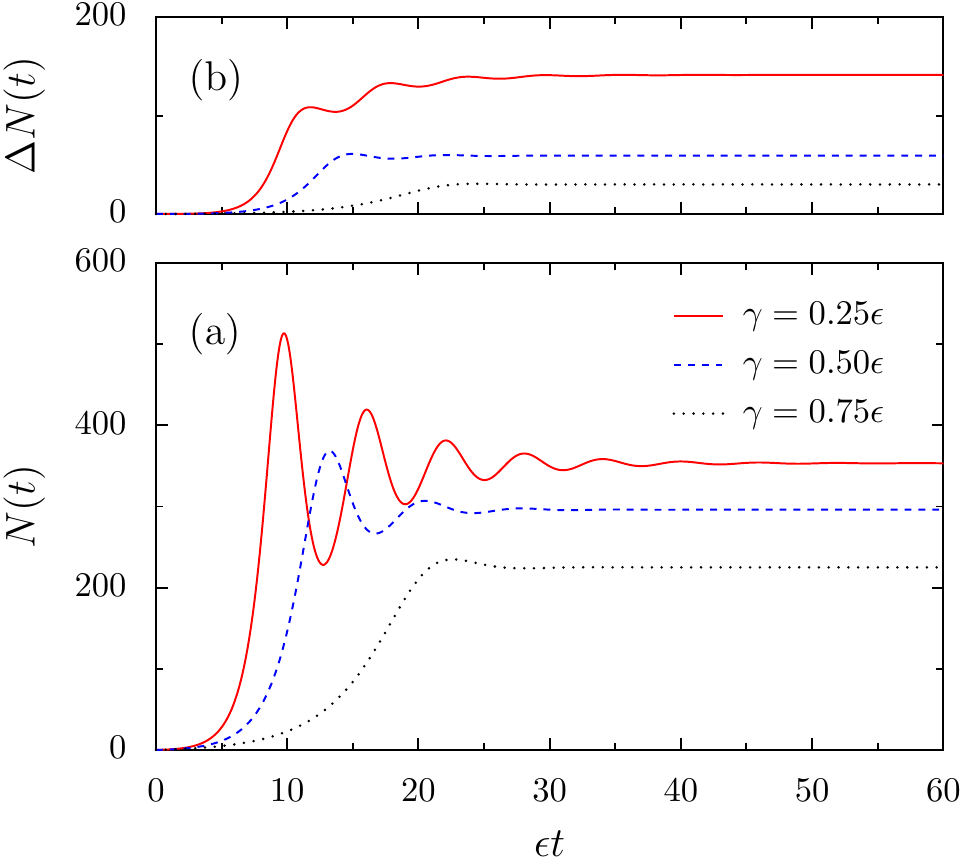}
\caption{\label{fig:anharmonic}%
(Color online) 
(a) Total number $N(t)$ and (b) difference $\Delta N(t)$ of bosonic excitations 
as a function of time from a numerical solution of
Eq.~\eqref{eq:eom_mean_field_RWA} at resonance ($\Omega=\omega_++\omega_-$) 
for various values of the damping constant $\gamma$.
In the figure, $T=0$,
$\Delta\gamma=-\epsilon/10$, $\omega_0=10^3\epsilon$, $\Omega=\sqrt{5}\omega_0$
(corresponding to $\omega_\mathrm{c}=\omega_0$), and $\alpha=10^{-6}$.}
\end{figure}

We now turn to the full solution of Eq.~\eqref{eq:eom_mean_field_RWA}, including
anharmonic terms, i.e., $\alpha\neq0$. To this end, we concentrate on the case
where the pumping frequency is at resonance, i.e., $\Omega=\omega_++\omega_-$.
A numerical solution of Eq.~\eqref{eq:eom_mean_field_RWA} is presented in
Fig.~\ref{fig:anharmonic} for the case where
$\epsilon>\sqrt{\gamma^2-\Delta\gamma^2}$, i.e., when parametric amplification takes
place in the absence of the anharmonic term ($\alpha=0$), see
Sec.~\ref{sec:damping}.
For short times, both the total number of bosonic modes $N(t)$
[Fig.~\ref{fig:anharmonic}(a)] and the difference between the occupation of the $+$ and $-$ modes,
$\Delta N(t)$ [Fig.~\ref{fig:anharmonic}(b)] get exponentially amplified. This
behavior is followed by a series of oscillations, to finally reach a stationary
occupation due to the residual (anharmonic) interaction. 

\begin{figure}[tb]
\includegraphics[width=\linewidth]{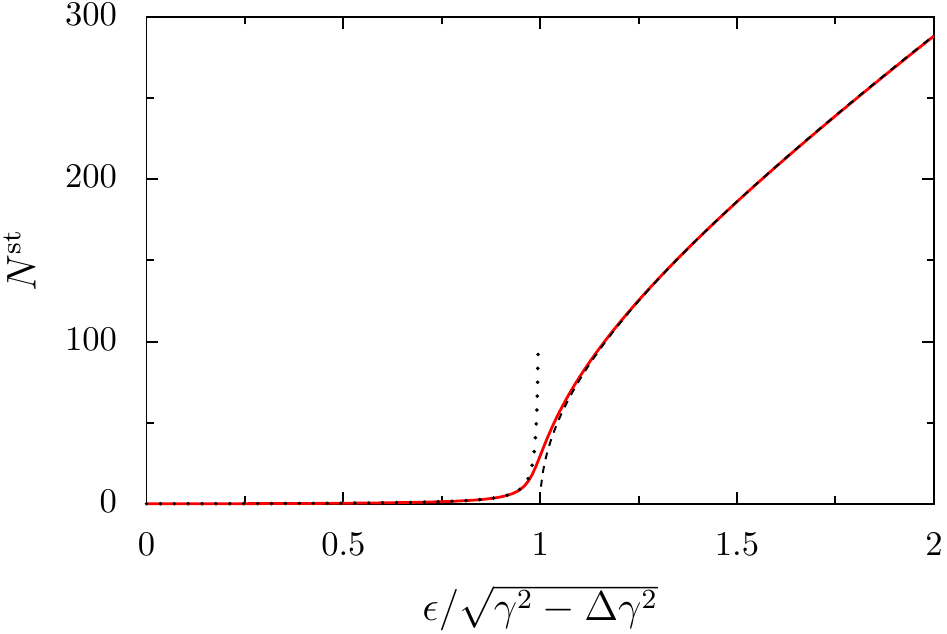}
\caption{\label{fig:stationary}%
(Color online) 
Stationary number of bosonic modes from Eq.~\eqref{eq:stationary} as a function
of the driving frequency $\epsilon$ (solid red line). The asymptotic behaviors
for $\epsilon<\sqrt{\gamma^2-\Delta\gamma^2}$ [Eq.~\eqref{eq:N_st_damping}] and
$\epsilon>\sqrt{\gamma^2-\Delta\gamma^2}$ [Eq.~\eqref{eq:N_st_amplified}] are
shown as a dotted and a dashed line, respectively. In the figure,
$\alpha=10^{-6}$ and $\omega_0/\gamma=2\times10^3$.}
\end{figure}

To better understand the numerical results of Fig.~\ref{fig:anharmonic}, we now
search for the stationary solution of Eq.~\eqref{eq:eom_mean_field_RWA}. Since
we work to first order in the small parameters $\epsilon/\Omega\ll1$ and
$\alpha\ll1$, we neglect the terms $\propto\alpha\epsilon/\Omega$ in
Eq.~\eqref{eq:eom_mean_field_RWA}, yielding 
\begin{align}
\label{eq:stationary}
\left(1-\frac{\Delta\gamma^2}{\gamma^2}\right)&\left(3\alpha\omega_0\right)^2
N^\mathrm{st}\left(N^\mathrm{st}+1\right)^2
\nonumber\\
&-\left(\epsilon^2+\Delta\gamma^2-\gamma^2\right)N^\mathrm{st}
-\epsilon^2=0
\end{align}
and
\begin{subequations}
\begin{align}
\label{eq:Delta_N_st_amplified}
\Delta N^\mathrm{st}&=-\frac{\Delta\gamma}{\gamma}N^\mathrm{st}, 
\\
\tilde\varphi^\mathrm{st}&=\frac{\epsilon(N^\mathrm{st}+1)/2}{\gamma+3\mathrm{i}\alpha\omega_0(N^\mathrm{st}+1)}.
\end{align}
\end{subequations}
The solution to Eq.~\eqref{eq:stationary} is shown in Fig.~\ref{fig:stationary}
as a red solid line as a function of the driving frequency $\epsilon$. 
For $\epsilon>\sqrt{\gamma^2-\Delta\gamma^2}$ and $\alpha=0$, we know from our study in
Sec.~\ref{sec:damping} that $N(t)$ diverges exponentially [cf.\
Eq.~\eqref{eq:damping_exp_N}], such that we can safely assume for that case
$N^\mathrm{st}\gg1$ in Eq.~\eqref{eq:stationary}. If on top of that,
$(\epsilon^2+\Delta\gamma^2-\gamma^2)N^\mathrm{st}\gg\epsilon^2$, we obtain
\begin{equation}
\label{eq:N_st_amplified}
N^\mathrm{st}\simeq\frac{\sqrt{\epsilon^2+\Delta\gamma^2-\gamma^2}}
{3\alpha\omega_0\sqrt{1-\Delta\gamma^2/\gamma^2}}, 
\end{equation}
showing that the smaller $\alpha$, the more efficient is the
parametric amplification. The asymptotic behavior of
Eq.~\eqref{eq:N_st_amplified} is shown as a dashed line in
Fig.~\ref{fig:stationary} and compares well with the full solution of
Eq.~\eqref{eq:stationary} (solid red line). For the case
$\epsilon<\sqrt{\gamma^2-\Delta\gamma^2}$, we
find that parametric amplification does not take place, and that the stationary
number of bosonic modes is, to first order in $\alpha$, given by
Eq.~\eqref{eq:N_st_damping}, as shown in Fig.~\ref{fig:stationary} by a dotted
line.

\section{Experimental consequences}
\label{sec:experiments}
Before presenting the various possible experimental consequences of our
proposal, we start this section by
discussing how one can trigger the parametric resonance of the magnetoplasmon
collective modes in realistic quasi-two-dimensional
semiconductor quantum dots. For a constant average electron density in the dot,
the confinement frequency entering the single-particle potential \eqref{eq:U} is
given by \cite{reima02_RMP, koski97_PRL}
\begin{equation}
\label{eq:omega_0}
\omega_0=\frac{e}{\sqrt{4\pi\epsilon_0\epsilon_\mathrm{r}m^* r_\mathrm{s}^3
}N_\mathrm{e}^{1/4}}
\end{equation}
and can be tuned by the gate voltage forming the quantum dot through the number
$N_\mathrm{e}$ of electrons.
In Eq.\ \eqref{eq:omega_0}, $\epsilon_0$ is the
vacuum permittivity, $\epsilon_\mathrm{r}$ the relative static dielectric
constant of the considered material, $r_\mathrm{s}$ the Wigner-Seitz radius, and
$m^*$ the effective electron mass.
Using the values for a GaAs quantum dot ($\epsilon_\mathrm{r}=13$, $m^*=0.067 m$
with $m$ the bare electronic mass, $r_\mathrm{s}=1.5 a_\mathrm{B}^*$ with
$a_\mathrm{B}^*=4\pi\epsilon_0\epsilon_\mathrm{r}\hbar^2/m^*e^2$ the effective
Bohr radius), and assuming $N_\mathrm{e}=10^3$, we have $\omega_0=\unit[1.7]{THz}$
($\hbar\omega_0=\unit[1.1]{meV}$). In a static magnetic field of
$B_0=\unit[2]{T}$, the eigenfrequencies \eqref{eq:omega_pm} are given by
$\omega_+=\unit[5.7]{THz}$ and $\omega_-=\unit[0.5]{THz}$. The resonance
condition $\Omega=\omega_++\omega_-$ for parametric amplification to occur [see
Eq.\ \eqref{eq:Omega_pm}]
thus requires in this case that the
magnetic field is periodically modulated at the frequency 
$\Omega/2\pi=\unit[0.99]{THz}$. The ongoing efforts towards the production of
THz sources should allow one to attain such pumping frequencies in a near future.
\cite{fergu02_NatureMat, tonou07_NaturePhoton, dyako10_CRP}
As we have shown in Secs. \ref{sec:damping} and \ref{sec:anharmonic}, parametric
amplification only occurs if the driving frequency $\epsilon$ of Eq.\
\eqref{eq:epsilon} is larger than the linewidth $\gamma$ of the magnetoplasmons
(assumed here to be the same for both collective modes). With a pumping strength
$\eta=0.1$, we obtain $\epsilon=\unit[0.22]{THz}$. Assuming
$\gamma\simeq\omega_0/10$ which is the typical value encountered in experiments,
\cite{allen83_PRB} parametric
amplification should occur in this case at a rate given by
$\epsilon-\gamma=\unit[50]{GHz}$ [see Eq.\ \eqref{eq:damping_exp_N}].

In what follows, we propose several ways of detecting the parametric amplification of the
magnetoplasmons:
First, these modes decay, among various processes, by radiative
damping, emitting photons at the frequencies $\omega_+$ and $\omega_-$.
Triggering the parametric resonance of the modes with the only help of a modulated magnetic 
field should thus result in the spontaneous emission of photons at these
frequencies, \textit{without} using an external
light source, as usually employed in far infrared spectroscopy experiments. 

Second, as we have shown in Sec.\ \ref{sec:PR}, the average fluctuations of the
electronic center-of-mass coordinate are directly related to the total number of
bosonic excitations $N$ in the system, $\overline{\langle\delta R^2\rangle}\sim
N$. As the latter is exponentially amplified up to its stationary value
\eqref{eq:N_st_amplified}
due to the anharmonicity, this should result in an expansion of the whole
electronic cloud forming the quantum dot which can be measured by scanning
tunneling microscopy. \cite{malte03_PRL}

Another measurable quantity \cite{ooste98_PRL, schwa02_JAP} related to the
fluctuations of the center-of-mass coordinate is the magnetization of the
quantum dot. Indeed, the magnetization operator for the electronic system
described by the Hamiltonian \eqref{eq:H(t)_def} reads 
\begin{equation}
\label{eq:M}
\mathbf{M}=-\frac{e}{2m^*}\sum_{i=1}^{N_\mathrm{e}}
\mathbf{r}_i\times\left[\mathbf{p}_i+e\mathbf{A}(\mathbf{r}_i, t)\right].
\end{equation}
Using Eq.\ \eqref{eq:A} and introducing center-of-mass and relative coordinates
as in Sec.\ \ref{sec:model}, the magnetization operator separates into
$\mathbf{M}=\mathbf{M}_\mathrm{cm}+\mathbf{M}_\mathrm{rel}$, with 
\begin{equation}
\label{eq:M_cm}
\mathbf{M}_\mathrm{cm}=-\frac{e}{2m^*}\left(L_Z+\frac{N_\mathrm{e}eB(t)}{2}R^2\right)\mathbf{e}_z
\end{equation}
the contribution from the center-of-mass dynamics, and 
\begin{equation}
\mathbf{M}_\mathrm{rel}=-\frac{e}{2m^*}\sum_i\left(l_{z,i}'+\frac{eB(t)}{2}{r_i'}^2\right)\mathbf{e}_z
\end{equation}
the one from the relative coordinates. Since the latter is induced by a bath of fermionic
quasiparticles, the modulation of the magnetic field does not lead to their
parametric amplification. In fact, the only result of the modulation is a
small periodic variation of the relative magnetization 
that averages to zero due to the fast modulation. As a consequence, 
the average magnetization resulting from the relative coordinates is constant in
time and equals its value without parametric modulation, 
$\overline{\langle\mathbf{M}_\mathrm{rel}\rangle}$. 

In contrast, the magnetization \eqref{eq:M_cm} resulting from the center-of-mass dynamics can
be conveniently rewritten in terms of the bosonic operators \eqref{eq:b_pm}.
Taking its expectation value, and averaging over fastly oscillating terms, we
find, to first order in the parametric modulation strength $\eta$,
\begin{equation}
\overline{\langle\mathbf{M}_\mathrm{cm}\rangle}(t)=
-\mu_\mathrm{B}^*\mathbf{e}_z\left\{\Delta
N(t)+\frac{\omega_\mathrm{c}/2}{\sqrt{\omega_0^2+\omega_\mathrm{c}^2/4}}\left[N(t)+1\right]\right\},
\end{equation}
with $\mu_\mathrm{B}^*=e\hbar/2m^*$ the effective Bohr magneton
($\mu_\mathrm{B}^*=\unit[0.86]{meV/T}$ for GaAs). As both the total number of
bosonic excitations $N(t)$ and the difference $\Delta N(t)$ are exponentially
amplified due to the parametric modulation to finally reach stationary values
[see Eqs.\ \eqref{eq:Delta_N_st_amplified} and \eqref{eq:N_st_amplified}], 
the resulting magnetization of the
dot should dramatically increase 
when the
parametric amplification of the magnetoplasmons takes place
as compared to its equilibrium value.

\section{Conclusion}
\label{sec:ccl}
In conclusion, we have studied the possibility of parametrically amplifying
bosonic collective modes in finite-size fermionic systems. Specifically,
we have shown that the magnetoplasmons in quasi-two-dimensional
semiconductor quantum dots can be parametrically amplified by modulating the
magnetic field perpendicular to the nanostructure. Moreover, we have
demonstrated that damping mechanisms and anharmonicities of the confinement lead
to a saturation of the parametric resonance. 
We have further discussed the implementation of our proposal in realistic
experimental samples and
suggested measurements that should present clear signatures of the parametric
amplification of the magnetoplasmons.

Our predictions
could in principle also apply to the magnetoplasmon modes in metallic
nanoparticles. However, as the damping of these modes is much
stronger than in quantum dots, \cite{weick11_PRB} it may be much more difficult to trigger the
parametric amplification in metallic nanoparticles.

\begin{acknowledgments}
We are particularly indebted to Rodolfo Jalabert for a discussion that inspired
this work and for his many useful comments and suggestions.
We would also like to thank Matthieu Bailleul, Bill Barnes, Stéphane Berciaud, Cosimo Gorini, Gert
Ingold, Pablo Tamborenea
and Dietmar Weinmann for discussions.
\end{acknowledgments}

\appendix
\section{Parametric resonance of a single harmonic oscillator: classical vs.\
quantum treatment}
\label{sec:PR_HO}
In this Appendix, we consider a single harmonic oscillator of mass $m$ whose
frequency $\omega(t)=\omega_0[1+\eta\sin{(\Omega t)}]$ is periodically modulated
in time at the frequency $\Omega$. Denoting $q$ and $p$ the position and the momentum of
the oscillator, its Hamiltonian reads
\begin{equation}
\label{eq:H_osc}
H_\mathrm{osc}(t)=\frac{p^2}{2m}+\frac{m}{2}\omega^2(t)q^2.
\end{equation}
In what follows, we assume that the strength $\eta$ of the periodic modulation
is much smaller than one, allowing for a perturbative treatment of the Hamiltonian
\eqref{eq:H_osc}.
For an arbitrary $\eta$, the Floquet formalism would be the appropriate
one. \cite{dittrich}
Before we tackle the quantum-mechanical treatment of the Hamiltonian
\eqref{eq:H_osc}, we recall in the next section the classical case.

\subsection{Classical treatment}
In order to find the classical trajectories associated to the Hamiltonian
\eqref{eq:H_osc}, as well as to stress the analogy with the quantum mechanical
treatment of the parametric resonance presented in the Sec.\ \ref{sec:HO_quantum}, it is
convenient to introduce the new variables
\begin{subequations}
\label{eq:z}
\begin{align}
z&=\frac{1}{\sqrt{2\mathrm{i}}}\left(\sqrt{m\omega_0}q+\frac{\mathrm{i}p}{\sqrt{m\omega_0}}\right),
\\
\bar
z&=\frac{1}{\sqrt{2\mathrm{i}}}\left(\sqrt{m\omega_0}q-\frac{\mathrm{i}p}{\sqrt{m\omega_0}}\right).
\end{align}
\end{subequations}
Since $z$ and $\bar z$ are canonically conjugated, i.e., the Poisson bracket
$\{\bar z, z\}=1$, the Hamilton equations of motion read 
$\dot z=-\partial H_\mathrm{osc}/\partial \bar z$ and 
$\dot{\bar z}=\partial H_\mathrm{osc}/\partial z$. With Eq.\ \eqref{eq:z}, the
Hamiltonian \eqref{eq:H_osc} transforms into
\begin{equation}
H_\mathrm{osc}(t)=\mathrm{i}\omega_0\left[1+\eta\sin{(\Omega t)}\right]\bar z z
+\mathrm{i}\eta\frac{\omega_0}{2}\sin{(\Omega t)}\left(\bar z^2+z^2\right), 
\end{equation}
yielding
\begin{equation}
\label{eq:dot_z}
\dot z=-\mathrm{i}\omega_0\left[1+\eta\sin{(\Omega t)}\right]z
-\mathrm{i}\omega_0\eta\sin{(\Omega t)}\bar z.
\end{equation}
Introducing 
\begin{equation}
z(t)=\exp{\left(-\mathrm{i}\omega_0\int_0^t\mathrm{d}s\left[1+\eta\sin{(\Omega
s)}\right]\right)}\mathcal{Z}(t), 
\end{equation}
Eq.\ \eqref{eq:dot_z} becomes, to first order in $\eta$, 
\begin{equation}
\label{eq:z_tilde}
\dot{\mathcal{Z}}=-\mathrm{i}\omega_0\eta\sin{(\Omega t)}\mathrm{e}^{2\mathrm{i}\omega_0 t}
\bar{\mathcal{Z}}.
\end{equation}
Close to the resonance condition $\Omega\simeq 2\omega_0$, we keep 
only the dominant term $\propto
\mathrm{e}^{-\mathrm{i}(\Omega-2\omega_0)t}$ such that Eq.\ \eqref{eq:z_tilde}
decouples and yields
\begin{equation}
\ddot{\mathcal{Z}}+\mathrm{i}(\Omega-2\omega_0)\dot{\mathcal{Z}
}+\left(\frac{\eta\omega_0}{2}\right)^2\mathcal{Z}=0.
\end{equation}
With the initial conditions $\mathcal{Z}(0)=z(0)$ and
$\dot{\mathcal{Z}}(0)=\eta\omega_0\bar{\mathcal{Z}}(0)/2$, we finally obtain 
\begin{align}
z(t)=&\;\frac{\mathrm{e}^{-\mathrm{i}\omega_0t}}{\Omega_--\Omega_+}
\bigg[
\left(\Omega_-\mathrm{e}^{\mathrm{i}\Omega_+t}-\Omega_+\mathrm{e}^{\mathrm{i}\Omega_-t}\right)z(0)
\nonumber\\
&+\mathrm{i}\frac{\eta\omega_0}{2}
\left(\mathrm{e}^{\mathrm{i}\Omega_+t}-\mathrm{e}^{\mathrm{i}\Omega_-t}\right)z^*(0)
\bigg],
\end{align}
with 
\begin{equation}
\label{eq:Omega_pm_osc}
\Omega_\pm=\omega_0-\frac{\Omega}{2}\pm\frac{1}{2}\sqrt{(\Omega-2\omega_0)^2-(\eta\omega_0)^2}.
\end{equation}
Thus, when $|\Omega-2\omega_0|<\eta\omega_0$, the frequencies $\Omega_\pm$
become imaginary, which leads to the exponential amplification of the motion of
the oscillator. Indeed, coming back to the original variables of the Hamiltonian
\eqref{eq:H_osc} and assuming $\Omega=2\omega_0$, one obtains
\begin{subequations}
\begin{align}
q(t)&=q(0)\cos{(\omega_0 t)}\mathrm{e}^{\eta\omega_0 t/2}
+\frac{p(0)}{m\omega_0}\sin{(\omega_0 t)}\mathrm{e}^{-\eta\omega_0 t/2},
\\
p(t)&=p(0)\cos{(\omega_0 t)}\mathrm{e}^{-\eta\omega_0 t/2}
-m\omega_0q(0)\sin{(\omega_0 t)}\mathrm{e}^{\eta\omega_0 t/2}.
\end{align}
\end{subequations}
Notice that the parametric amplification only takes place within the above
approximate treatment if initially, $q(0)\neq0$.

\subsection{Quantum treatment}
\label{sec:HO_quantum}
At the quantum level, the Hamiltonian \eqref{eq:H_osc} is conveniently rewritten
as 
\begin{align}
H_\mathrm{osc}(t)=&\;\hbar\omega_0\left[1+\eta\sin{(\Omega t)}\right]
\left(b^\dagger b+\frac 12\right)
\nonumber\\
&+\eta\frac{\hbar\omega_0}{2}\sin{(\Omega t)}\left({b^\dagger}^2+b^2\right)
\end{align}
in terms of the lowering and raising bosonic operators
\begin{subequations}
\begin{align}
b=&\frac{1}{\sqrt{2}}\left(\frac{q}{\ell_\mathrm{osc}}+\frac{\mathrm{i}p\ell_\mathrm{osc}}{\hbar}\right),
\\
b^\dagger=&\frac{1}{\sqrt{2}}\left(\frac{q}{\ell_\mathrm{osc}}-\frac{\mathrm{i}p\ell_\mathrm{osc}}{\hbar}\right),
\end{align}
\end{subequations}
respectively. Here, $\ell_\mathrm{osc}=\sqrt{\hbar/m\omega_0}$ denotes the
oscillator length.
The Heisenberg equation of motion for the operator $b$ reads
\begin{equation}
\dot b=-\mathrm{i}\omega_0\left[1+\eta\sin{(\Omega t)}\right]b
-\mathrm{i}\omega_0\eta\sin{(\Omega t)}b^\dagger 
\end{equation}
and is similar to the classical equation of motion \eqref{eq:dot_z} for the
classical variable $z$. It can thus be solved in exactly the same way, yielding
\begin{align}
b(t)=&\;\frac{\mathrm{e}^{-\mathrm{i}\omega_0t}}{\Omega_--\Omega_+}
\bigg[
\left(\Omega_-\mathrm{e}^{\mathrm{i}\Omega_+t}-\Omega_+\mathrm{e}^{\mathrm{i}\Omega_-t}\right)b(0)
\nonumber\\
&+\mathrm{i}\frac{\eta\omega_0}{2}
\left(\mathrm{e}^{\mathrm{i}\Omega_+t}-\mathrm{e}^{\mathrm{i}\Omega_-t}\right)b^\dagger(0)
\bigg],
\end{align}
with the frequencies $\Omega_\pm$ given in Eq.\ \eqref{eq:Omega_pm_osc}. Thus,
at the quantum-mechanical level, the lowering and raising operators are
exponentially amplified whenever the resonance condition
$|\Omega-2\omega_0|<\eta\omega_0$ is fulfilled.

Considering that the harmonic oscillator is in its ground state before the parametric
amplification takes place ($t<0$), and assuming $\Omega=2\omega_0$, 
the average number of bosonic excitations 
\begin{equation}
\langle b^\dagger b \rangle(t)=\frac{1}{2}\left[\cosh{(\eta\omega_0 t)}-1\right]
\end{equation}
exponentially increases for increasing time.
During that process, the average
position and momentum are both vanishing, $\langle q \rangle(t)=\langle p
\rangle(t)=0$, while the corresponding fluctuations are exponentially amplified, 
\begin{subequations}
\begin{align}
\langle \delta q^2 \rangle(t)=\frac{\ell_\mathrm{osc}^2}{2}
\left[\cosh{(\eta\omega_0 t)}+\cos{(2\omega_0 t)}\sinh{(\eta\omega_0 t)}\right],
\\
\langle \delta p^2 \rangle(t)=\frac{\hbar^2}{2\ell_\mathrm{osc}^2}
\left[\cosh{(\eta\omega_0 t)}-\cos{(2\omega_0 t)}\sinh{(\eta\omega_0 t)}\right].
\end{align}
\end{subequations}
The quantum fluctuations of the oscillator in its ground state thus provide a
seed for the parametric resonance in the quantum case.

\section{Parametric resonance of the magnetoplasmons: classical treatment}
\label{sec:PR_classical}
Classically, the center of mass Hamiltonian \eqref{eq:H_cm(t)_coordinates} in
the absence of the quartic correction ($A_4=0$) can
be solved by a series of two successive canonical transformations. First, we
introduce
\begin{subequations}
\begin{align}
\xi&=\frac{1}{\sqrt{2}}(X+\mathrm{i}Y), \qquad
P_\xi=\frac{1}{\sqrt{2}}(P_X-\mathrm{i}Y),\\
\xi^*&=\frac{1}{\sqrt{2}}(X-\mathrm{i}Y), \qquad
P_{\xi^*}=\frac{1}{\sqrt{2}}(P_X+\mathrm{i}Y),
\end{align}
\end{subequations}
with Poisson brackets $\{\xi, P_\xi\}=\{\xi^*, P_{\xi^*}\}=1$,
such that the Hamiltonian \eqref{eq:H_cm(t)_coordinates} transforms into
\begin{align}
\label{eq:H_xi}
H_\mathrm{cm}(t)=&\;\frac{P_\xi P_{\xi^*}}{M}
+M\left(\omega_0^2+\frac{\omega^2(t)}{4}\right)\xi\xi^*
\nonumber\\
&+\mathrm{i}\frac{\omega(t)}{2}\left(\xi P_\xi-\xi^* P_{\xi^*}\right).
\end{align}
Second, we introduce the canonically-conjugated variables
\begin{subequations}
\begin{align}
z_+&=\frac{1}{\sqrt{2\mathrm{i}}}\left(\beta\xi^*+\frac{\mathrm{i}P_\xi}{\beta}\right),
\quad
\bar z_+=\frac{1}{\sqrt{2\mathrm{i}}}\left(\beta\xi-\frac{\mathrm{i}P_{\xi^*}}{\beta}\right),
\\
z_-&=\frac{1}{\sqrt{2\mathrm{i}}}\left(\beta\xi+\frac{\mathrm{i}P_{\xi^*}}{\beta}\right),
\quad
\bar z_-=\frac{1}{\sqrt{2\mathrm{i}}}\left(\beta\xi^*-\frac{\mathrm{i}P_\xi}{\beta}\right),
\end{align}
\end{subequations}
with $\beta=\sqrt{M(\omega_0^2+\omega_\mathrm{c}^4/4)^{1/2}}$ and the Poisson
brackets $\{\bar z_\pm, z_\pm\}=1$. In terms of these variables, the Hamiltonian
\eqref{eq:H_xi} reads
\begin{align}
\label{eq:H_z_pm}
H_\mathrm{cm}(t)=&\;\mathrm{i}\sum_{\sigma=\pm}\omega_\sigma\left[1+\sigma\delta f(t)\right]\bar
z_\sigma z_\sigma
\nonumber\\
&+\mathrm{i}\frac{\omega_\mathrm{c}}{2}\delta f(t)\left(\bar z_+\bar
z_-+z_+z_-\right), 
\end{align}
where the function $\delta f(t)$ is defined in Eq.\ \eqref{eq:delta_f} and the
frequencies $\omega_\pm$ are given in Eq.\ \eqref{eq:omega_pm}.
The Hamilton equations of motion associated to the Hamiltonian \eqref{eq:H_z_pm}
thus read
\begin{equation}
\dot z_\pm=-\frac{\partial H_\mathrm{cm}}{\partial \bar z_\pm}
=-\mathrm{i}\omega_\pm\left[1\pm\delta f(t)\right]z_\pm
-\mathrm{i}\frac{\omega_\mathrm{c}}{2}\delta f(t)\bar z_\mp.
\end{equation}
Being similar to the quantum mechanical Heisenberg equations of motion
\eqref{eq:eom_b_pm} for the bosonic operators $b_\pm$, the above equation can be
solved in exactly the same way as in Sec.\ \ref{sec:PR}. Going back to the
original variables of the center of mass Hamiltonian
\eqref{eq:H_cm(t)_coordinates}, we find that close to resonance
[$|\Omega-\omega_+-\omega_-|\ll\epsilon$ with $\epsilon$ given in Eq.\
\eqref{eq:epsilon}], the classical motion of the center of mass is described by
the trajectories
\begin{subequations}
\label{eq:trajectories}
\begin{align}
X(t)=&\;\left\{X(0)\left[\cos{(\omega_+ t)}+\cos{(\omega_- t)}\right]\right.
\nonumber\\
&\left.-Y(0)\left[\sin{(\omega_+ t)}-\sin{(\omega_- t)}\right]
\right\}\frac{\mathrm{e}^{\epsilon t/2}}{2}
\nonumber\\
&+
\left\{P_X(0)\left[\sin{(\omega_+ t)}+\sin{(\omega_- t)}\right]\right.
\nonumber\\
&\left.+P_Y(0)\left[\cos{(\omega_+ t)}-\cos{(\omega_- t)}\right]
\right\}\frac{\mathrm{e}^{-\epsilon t/2}}{2\beta^2}, 
\\
Y(t)=&\;\left\{Y(0)\left[\cos{(\omega_+ t)}+\cos{(\omega_- t)}\right]\right.
\nonumber\\
&\left.+X(0)\left[\sin{(\omega_+ t)}-\sin{(\omega_- t)}\right]
\right\}\frac{\mathrm{e}^{\epsilon t/2}}{2}
\nonumber\\
&+
\left\{P_Y(0)\left[\sin{(\omega_+ t)}+\sin{(\omega_- t)}\right]\right.
\nonumber\\
&\left.-P_X(0)\left[\cos{(\omega_+ t)}-\cos{(\omega_- t)}\right]
\right\}\frac{\mathrm{e}^{-\epsilon t/2}}{2\beta^2}.
\end{align}
\end{subequations}

\begin{figure*}[tb]
\includegraphics[width=\linewidth]{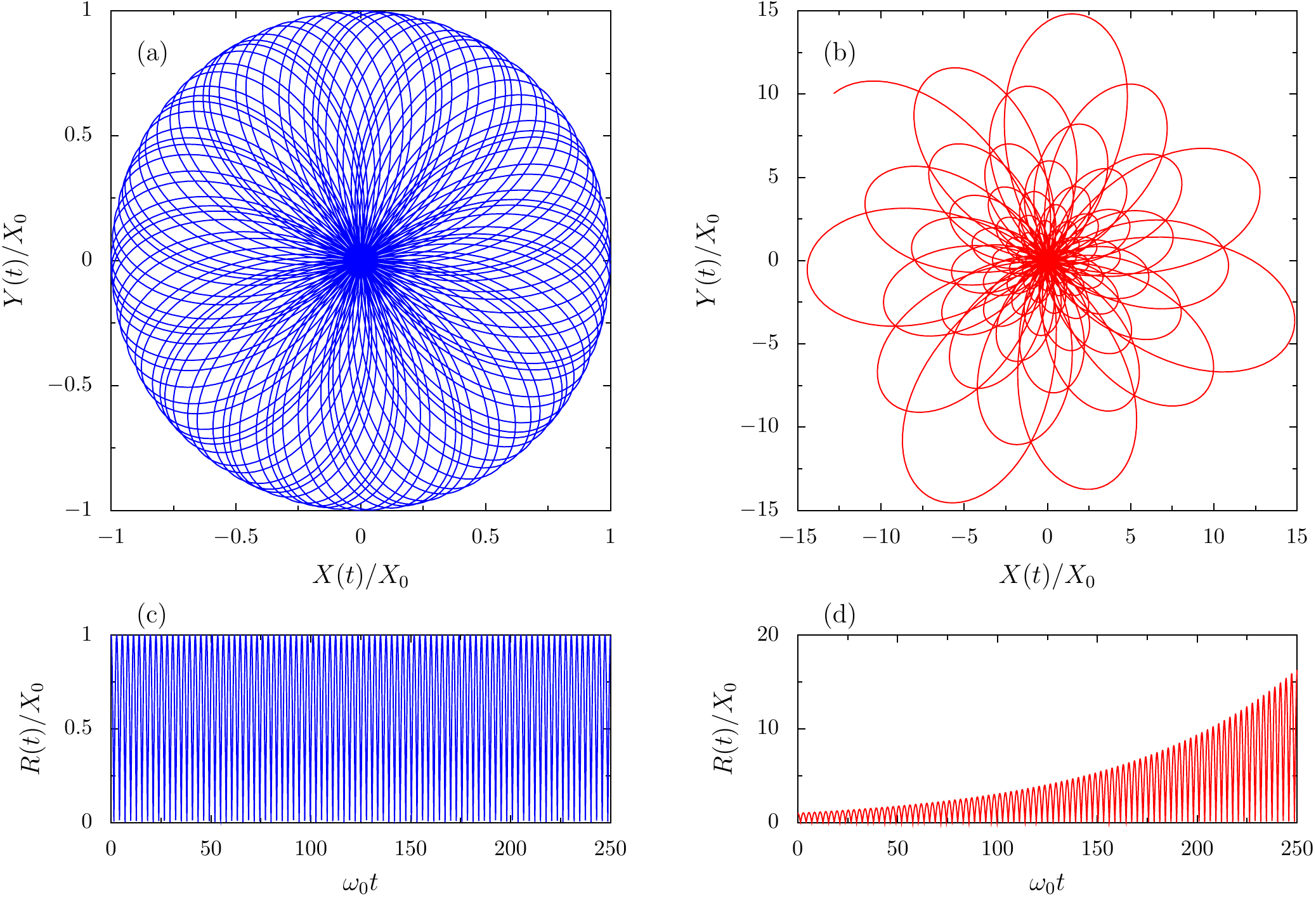}
\caption{\label{fig:trajectories}%
(Color online) 
Upper panel: Typical trajectory $\big(X(t), Y(t)\big)$ of the center of mass coordinate (a) in the
absence of parametric modulation ($\eta=0$) and (b) with parametric
modulation ($\eta=0.1$) for times up to $t_\mathrm{max}=250/\omega_0$. Lower panel: radial coordinate
$R(t)=\sqrt{X^2(t)+Y^2(t)}$ as a function of time for (c) $\eta=0$ and (d)
$\eta=0.1$. For all trajectories, the initial conditions are $X(0)=X_0$, $Y(0)=0$
and $P_X(0)=P_Y(0)=0$. In the figure, the cyclotron frequency
$\omega_\mathrm{c}=\omega_0$ and the pumping frequency is at resonance, i.e.,
$\Omega=\omega_++\omega_-$.}
\end{figure*}

In Fig.\ \ref{fig:trajectories}, we show typical trajectories of the center of
mass from Eq.\ \eqref{eq:trajectories} in absence of parametric modulation [Fig.\ \ref{fig:trajectories}(a)] 
and with parametric modulation [Fig.\ \ref{fig:trajectories}(b)]. Without
parametric modulation, the center of mass has trajectories corresponding to a
two-dimensional harmonic oscillator in a perpendicular magnetic field [Fig.\ \ref{fig:trajectories}(a)]. 
The corresponding radial coordinate shown in Fig.\ \ref{fig:trajectories}(c)
thus regularly oscillates between $0$ and its maximal amplitude. In contrast, in
the presence of parametric modulation [Figs.\ \ref{fig:trajectories}(b) and
\ref{fig:trajectories}(d)], the motion of the center of mass is exponentially
amplified.

\section{Toy model: Parametric modulation of two coupled fermionic modes}
\label{sec:toy}
In this Appendix, we consider two coupled fermionic modes with frequencies
$\epsilon_+$ and $\epsilon_-$ whose
corresponding Hamiltonian is parametrically modulated. In the sequel, we show
that due to the fermionic nature of the modes, Pauli principle prevents the
parametric amplification to occur.

As a toy model, we consider a Hamiltonian similar to Eq. \eqref{eq:H_cm(t)}
without quartic correction, except that the bosonic operators in Eq.\
\eqref{eq:H_cm(t)} are replaced by fermionic ones,
\begin{align}
\label{eq:H_toy}
H_\mathrm{toy}(t)=&
\sum_{\sigma=\pm}\hbar \epsilon_\sigma\left[1+\sigma\eta\sin{(\Omega t)}\right]
c_\sigma^\dagger c^{\phantom{\dagger}}_\sigma
\nonumber\\
&+\eta\frac{\hbar\Delta}{2}\sin{(\Omega t)}\left(c_+^\dagger
c_-^\dagger+c_-c_+\right).
\end{align}
Here, $c_\sigma^\dagger$/$c_\sigma$ creates/annihilates a fermion in mode
$\sigma=\pm$. In Eq.\ \eqref{eq:H_toy}, $\eta\ll1$ is the strength of the
parametric modulation at frequency $\Omega$ and $\Delta$ is the frequency scale of
a broken symmetry term. The corresponding Heisenberg equations of motion read
\begin{equation}
\label{eq:eom_c}
\dot c_\pm=
-\mathrm{i}\epsilon_\pm\left[1\pm\eta\sin{(\Omega t)}\right]c_\pm
\mp\mathrm{i}\eta\frac{\Delta}{2}\sin{(\Omega t)}c_\mp^\dagger.
\end{equation}

Along the same lines as in Sec.\ \ref{sec:PR}, introducing the new operators
\begin{equation}
c_\pm(t)=\exp{\left(-\mathrm{i}\epsilon_\pm
\int_0^t\mathrm{d}s\left[1\pm\eta\sin{(\Omega s)}\right]\right)}\tilde c_\pm(t), 
\end{equation}
we obtain for Eq.\ \eqref{eq:eom_c} to first order in $\eta$ and within the
rotating wave approximation,
which is valid for $\Omega\simeq\epsilon_++\epsilon_-$,
\begin{equation}
\ddot{\tilde c}_\pm+\mathrm{i}\left(\Omega-\epsilon_+-\epsilon_-\right)\dot{\tilde c}_\pm
+\left(\frac{\eta\Delta}{4}\right)^2\tilde c_\pm=0.
\end{equation}
Solving for the latter equation leads to 
\begin{align}
\label{eq:full_solution_c}
c_\pm(t)=&\;\frac{\mathrm{e}^{-\mathrm{i}\epsilon_\pm t/\hbar}}{\Omega_--\Omega_+}
\bigg[
\left(\Omega_-\mathrm{e}^{\mathrm{i}\Omega_+t}-\Omega_+\mathrm{e}^{\mathrm{i}\Omega_-t}\right)c_\pm(0)
\nonumber\\
&\pm\mathrm{i}\frac{\eta\Delta}{4}
\left(\mathrm{e}^{\mathrm{i}\Omega_+t}-\mathrm{e}^{\mathrm{i}\Omega_-t}\right)c_\mp^\dagger(0)
\bigg],
\end{align}
with 
\begin{equation}
\Omega_\pm=\frac{\epsilon_++\epsilon_--\Omega}{2}\pm\frac 12
\sqrt{(\Omega-\epsilon_+-\epsilon_-)^2+\frac{(\eta\Delta)^2}{4}}.
\end{equation}
Thus, contrarily to the bosonic case [cf.\ Eq.\ \eqref{eq:Omega_pm}], the
frequencies $\Omega_\pm$ are real for any pumping frequency $\Omega$, such that
the fermionic operators \eqref{eq:full_solution_c} are not parametrically
amplified. This is a direct consequence of Pauli's principle.



\begin{thebibliography}{99}
\bibitem{footnote:Jacobi}
We assumed for obtaining the decomposition~\eqref{eq:H(t)_separation} that the number 
of electrons $N_\mathrm{e}$ is sufficiently large such that
$N_\mathrm{e}\simeq N_\mathrm{e}-1$. Thus, the number of relative degrees of freedom equals the total
one. Notice that the decomposition~\eqref{eq:H(t)_separation} can be more
rigorously obtained by means of Jacobi coordinates (see, e.g.,
Ref.~\onlinecite{jacak}), 
which would leave the center-of-mass Hamiltonian~\eqref{eq:H_cm(t)_coordinates} unchanged.
\end{thebibliography}
\end{document}